# Distributionally Robust Evaluation for Real-Time Flexibility of Electric Vehicles Considering Uncertain Departure Behavior and State-of-Charge

Yixin Li, *Student Member, IEEE,* Zhengshuo Li, *Senior Member, IEEE*

*Abstract*—Accurately evaluating the real-time flexibility of electric vehicles (EVs) is necessary for EV aggregators to offer ancillary services. However, regulation-caused uncertain state-of-charge and random departure behavior complicate the evaluation and badly impact the evaluation accuracy. To resolve this issue, this letter proposes a distributionally robust real-time flexibility evaluation model that formulates the uncertain departure behavior and state-of-charge of EVs in an online updating pattern. Thanks to dualization, this model can be efficiently solved via off-the-shelf solvers. Case studies validate the superiority of the proposed method and its scalability regarding EV numbers.

## I. Introduction

WITH large-scale renewable energy integrated into power systems, demand-side resources, e.g., electric vehicles (EVs), are allowed to participate in real-time markets to provide frequency regulation services for system operators (SOs) [1]. In this context, an EV aggregator, e.g., the entity that manages an array of EVs in a charging station, should accurately evaluate the real-time flexibility denoted by the power adjustment range [2] when bidding in the real-time market.

The full-day flexibility of EVs is studied in [3], and the online flexibility of EVs is investigated in [4]. Despite these studies, accurate evaluation of real-time flexibility is still challenging because of the uncertainty in EV behavior and parameters. For example, a grid-connected EV is likely to leave the station earlier than what was previously reported to the aggregator. This stochastic early departure behavior complicates real-time flexibility evaluation, but neglecting this factor may yield unreliable results. Another form of uncertainty comes from EVs' state of charge (SOC) values due to the temporal settings in the real-time market and regulation [5][6], which are explained in Section II. This factor greatly impacts the evaluation accuracy but is rarely considered in the relevant literature.

To this end, in this letter, a distributionally robust evaluation method for real-time flexibility is proposed considering the uncertainty in both EVs' SOC and departure time. To the best of the authors' knowledge, this is the ***first*** real-time flexibility evaluation work to explicitly incorporate the impact of EVs' participation in the market and regulation on EVs' SOC uncertainty. Moreover, a distributionally robust ambiguity set about uncertain departure behavior is built in a data-driven and online updating pattern. By the duality theorem, the resultant flexibility evaluation model is transformed into a mixed-integer linear programming (MILP) problem that can be efficiently solved via off-the shelf solvers. Numerical tests verify the improved accuracy of the proposed method and its scalability regarding EV numbers.

## II. Uncertainty Model

### A. Impact of Participating in Real-time Market and Regulation

To establish the uncertainty model regarding EVs' SOC values, the impact of EVs' participation in the market and regulation is first explained below.

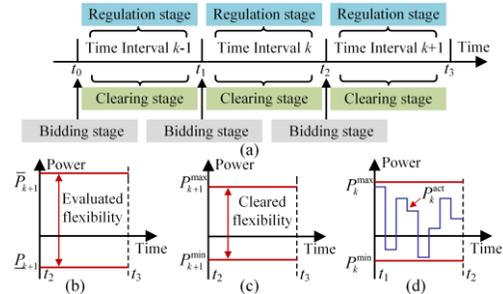

Fig. 1. The process of an aggregator participating in a real-time market to provide secondary frequency regulation. (a) The temporal settings; (b) The flexibility evaluated by the aggregator; (c) The flexibility cleared in the market; (d) The real-time regulation signal sent by the SO to the aggregator.

Fig. 1 briefly shows the process of an aggregator participating in a real-time market to provide secondary frequency regulation. The temporal settings in Fig. 1(a) include the bidding stage, clearing stage and regulation stage. In the *bidding* stage occurring at the start of each time interval (e.g., time $t_1$ for interval $k$), the aggregator must evaluate the flexibility for the next interval $k+1$ rather than the interval $k$ [5], as shown in Fig. 1(b), and form the bidding strategy. Here, $[\underline{P}_{k+1}, \overline{P}_{k+1}]$ is the ***evaluated*** flexibility. In the *clearing* stage of interval $k$, the market clearing result for interval $k+1$ is achieved, as shown in Fig. 1(c), where $[P_{k+1}^{\min}, P_{k+1}^{\max}]$ denotes the ***cleared*** flexibility. In the *regulation* stage of interval $k$, the aggregator receives the real-time regulation signal from the SO, e.g., the one depicted in Fig. 1(d), and it decomposes this signal among the EVs for implementation. Here, $P_k^{\text{act}}$ is the actual operating power of the aggregator affected by the real-time regulation signal, which is equal to the sum of the operating power of all EVs.

Notice that the above real-time regulation signal from the SO is issued based on the real-time frequency regulation requirement. Hence, the actual operating power $P_k^{\text{act}}$ of the aggregator, which determines the EVs' SOC at time $t_2$, ***cannot*** be known by the aggregator when the latter evaluates the flexibility at time $t_1$. In other words, the EVs' SOC at time $t_2$, denoted by $E_{k,n}$, is ***uncertain*** when the flexibility is evaluated

This work was supported by the National Natural Science Foundation of China under Grant 52377107.

Yixin Li and Zhengshuo Li are with the School of Electrical Engineering, Shandong University, Jinan 250061, China. Zhengshuo Li is the corresponding author (e-mail: zsli@sdu.edu.cn).

for the interval $k+1$. Since an EV's SOC value directly influences how much energy the EV can discharge or charge in the interval $k+1$, the above uncertainty must be considered in the flexibility evaluation.

*B. The Uncertainty Model of SOC*

The difficulty in modeling the above SOC uncertainty lies in estimating $P_k^{act}$. The aggregator at time $t_1$ knows only the $[\underline{P}_k^{min}, \overline{P}_k^{max}]$ for interval $k$ that was cleared in interval $k-1$, and it is *unaware* of the SO's regulation signal during interval $k$. As a result, an uncertainty interval is built to model the uncertainty $E_{k,n}$. Let $[E_{k,n}^{min}, E_{k,n}^{max}]$ denote the interval in which the SOC of the $n^{th}$ EV at time $t_2$ should reside. Either of its endpoints can be *online updated* (i.e., updated over time) as outlined below.

Take the $E_{k,n}^{min}$ evaluation as an example. As shown in Fig. 1(d), although a specific regulation trajectory is unknown, $P_k^{act}$ must reside in $[\underline{P}_k^{min}, \overline{P}_k^{max}]$ according to the current market rule. A straightforward idea is to consider the extreme case by setting $P_k^{act}$ equal to $\underline{P}_k^{min}$, and $E_{k,n}^{min}$ can thus be *safely* estimated as follows:

$$E_{k,n}^{min} = \arg\min_{p_{k,n}, E_{k,n}} \left\{ \sum_{n=1}^{N_k} E_{k,n} \left| \begin{array}{l} (p_{k,n}, E_{k,n}) \in \Phi_{k,n}, n \in N_k, \\ \underline{P}_k^{min} = \sum_{n=1}^{N_k} p_{k,n} \end{array} \right. \right\}, \quad (1)$$

where $N_k$ is the number set of EVs in interval $k$, $p_{k,n}$ is the $n^{th}$ EV's operating power in interval $k$, and $\Phi_{k,n}$ is the $n^{th}$ EV's feasible operation region in interval $k$, as formulated in

$$\Phi_{k,n} = \left\{ (p_{k,n}, E_{k,n}) \left| \begin{array}{l} 0 \leq p_{k,n}^{ch} \leq \overline{p}_n^{ch} z_{k,n}^{ch}, \ 0 \leq p_{k,n}^{dis} \leq \overline{p}_n^{dis} z_{k,n}^{dis}, \\ z_{k,n}^{ch} + z_{k,n}^{dis} \leq 1, \ p_{k,n} = p_{k,n}^{ch} - p_{k,n}^{dis}, \\ E_{k,n} = E_{k-1,n} + \eta_n^{ch} p_{k,n}^{ch} \Delta T - p_{k,n}^{dis} \Delta T / \eta_n^{dis}, \\ \underline{E}_{k,n} \leq E_{k,n} \leq \overline{E}_{k,n} \end{array} \right. \right\}, \quad (2)$$

where $p_{k,n}^{ch}, \overline{p}_n^{ch}, z_{k,n}^{ch}, \eta_n^{ch}$ are the charging power, power limit, binary variable and efficiency, $p_{k,n}^{dis}, \overline{p}_n^{dis}, z_{k,n}^{dis}, \eta_n^{dis}$ are the discharge-related parameters corresponding to the former, and $\overline{E}_{k,n}^{ev}, \underline{E}_{k,n}^{ev}$ are the SOC upper and lower limits of the $n^{th}$ EV at the end of interval $k$. $\Phi_{k,n}$ includes the operating power constraint, charging/discharge power constraint, SOC change process constraint, charge and discharge state constraint, and SOC limit constraint. More details can be found in [7].

Formula (1) ensures that $E_{k,n}^{min}$ is obtained by minimizing the sum of $E_{k,n}$ in the extreme case where $P_k^{act}$ equals $\underline{P}_k^{min}$. With $E_{k,n}^{max}$ being estimated similarly, the uncertainty interval model $U_{k,n} = \{E_{k,n} | E_{k,n}^{min} \leq E_{k,n} \leq E_{k,n}^{max}\}$ can be obtained.

*C. The Uncertainty Model of Unexpected EV Departure*

According to [8], EV users inform the aggregator of an estimated departure time when the EV arrives at the charging station. However, the EV actual departure time may not be consistent with the one previously reported, and the unexpected early departure usually results in a reduction in EVSC, which should be considered by the aggregator.

A naive idea is to directly model this impact in a bottom-up fashion, which requires establishing the statistical early departure model for every EV and then calculating the uncertain impact. Unfortunately, this avenue might be too complicated. Instead, in this letter, we propose modeling the proportion of unexpected-departure-caused-reduced flexibility to the total flexibility in a data-driven and online updating pattern.

The parameter $u$ is introduced to represent the proportion of the flexibility caused by the early departure of EVs. Assuming that the aggregator knows the EVs' estimated departure time and actual departure time, which is a weak and common assumption [8], then $u_m$ for the $m^{th}$ interval in history can be calculated by

$$u_m = (\overline{P}_m^{act} - \underline{P}_m^{act}) / (\overline{P}_m^0 - \underline{P}_m^0), \quad (3)$$

where $[\underline{P}_m^{act}, \overline{P}_m^{act}]$ represents the actual flexibility and $[\underline{P}_m^0, \overline{P}_m^0]$ represents the *virtual* flexibility as if no EVs had left unexpectedly. Then, $\underline{P}_m^{act}$ and $\underline{P}_m^0$ in (3) can be obtained by

$$\underline{P}_m^{act} = \min_{p_{m,n}, e_{m,n}} \left\{ \sum_{n=1}^{N_m} p_{m,n} \left| (p_{m,n}, e_{m,n}) \in \Phi_{m,n}, n \in N_m \right. \right\}, \quad (4)$$

$$\underline{P}_m^0 = \min_{p_{m,n}, e_{m,n}} \left\{ \sum_{n=1}^{N_m^0} p_{m,n} \left| (p_{m,n}, e_{m,n}) \in \Phi_{m,n}, n \in N_m^0 \right. \right\}, \quad (5)$$

where $N_m$ represents the actual set of EVs and $N_m^0$ represents the set of EVs if no EV leaves early. $\overline{P}_m^{act}$ and $\overline{P}_m^0$ can be obtained similarly. Thus, $u_m$ for historical intervals can be obtained.

After the sample set $\hat{\Xi} = \{\hat{u}_1, \hat{u}_2, ..., \hat{u}_i, ..., \hat{u}_S\}$ is calculated by the above method, the empirical distribution of $u$ can be established by $\hat{\mathbb{P}} := 1/S \sum_{i=1}^{S} \delta_{\hat{u}_i}$. Based on $\hat{\mathbb{P}}$, the ambiguity set $\mathcal{D}$ can be denoted by

$$\mathcal{D} := \{\mathbb{P} \in \mathcal{P}(\Xi) : W(\mathbb{P}, \hat{\mathbb{P}}) \leq \varepsilon\}, \quad (6)$$

in a *data-driven and online updating* pattern (i.e., the sample set is updated over time with new data). Here, $\mathcal{P}(\Xi)$ is the set of all distributions on the polyhedron $\Xi = \{u \in \mathbb{R} : Hu \leq h\}$, where $h = [\min\{\hat{u}_i\}, \max\{\hat{u}_i\}]^T$ and $H = [1, 1]^T$, $W$ is the Wasserstein distance, and $\varepsilon$ is the Wasserstein ball radius.

## III. DISTRIBUTIONALLY ROBUST EVALUATION MODEL

With the uncertainty set $U_{k,n}$ and ambiguity set $\mathcal{D}$, a distributionally robust evaluation model for $\underline{P}_{k+1}$ is formulated as follows:

$$\underline{P}_{k+1} = \min_{p_{k+1,n}, E_{k+1,n}} \max_{E_{k,n} \in U_{k,n}, \mathbb{P} \in \mathcal{D}} \mathbb{E}_{\mathbb{P}} \left\{ u \sum_{n=1}^{N_{k+1}} p_{k+1,n} \right\} \quad (7)$$
$$\text{s.t.} \quad (p_{k+1,n}, E_{k+1,n}) \in \Phi_{k+1,n}, n \in N_{k+1}$$

The model for $\overline{P}_{k+1}$ can be conducted similarly, so that description is omitted here.

By the duality theorem, (7) can be converted into an MILP problem as follows:

$$\begin{aligned}
\underline{P}_{k+1} = \ & \min \quad \eta \varepsilon + 1/S \times \sum_{i=1}^{S} \beta_i \\
\text{s.t.} \ & 0 \leq p_{k+1,n}^{ch} \leq \overline{p}_n^{ch} z_{k+1,n}^{ch}, \ 0 \leq p_{k+1,n}^{dis} \leq \overline{p}_n^{dis} z_{k+1,n}^{dis}, \ n \in N_{k+1}, \\
& z_{k+1,n}^{ch} + z_{k+1,n}^{dis} \leq 1, \ p_{k+1,n} = p_{k+1,n}^{ch} - p_{k+1,n}^{dis}, \ n \in N_{k+1}, \\
& -(E_{k,n}^{max} + E_{k,n}^{min})/2 + q_{1,n} - \eta_n^{ch} p_{k+1,n}^{ch} \Delta T \\
& + p_{k+1,n}^{dis} \Delta T / \eta_n^{dis} + \underline{E}_{k+1,n} + Z_{1,n} \Gamma \leq 0, \ n \in N_{k+1}, \\
& (E_{k,n}^{max} + E_{k,n}^{min})/2 + q_{2,n} + \eta_n^{ch} p_{k+1,n}^{ch} \Delta T \\
& - p_{k+1,n}^{dis} \Delta T / \eta_n^{dis} - \overline{E}_{k+1,n} + Z_{2,n} \Gamma \leq 0, \ n \in N_{k+1}, \\
& Z_{1,n} + q_{1,n} \geq (E_{k,n}^{max} - E_{k,n}^{min})/2 \times y_1, \ q_{1,n} \geq 0, \ Z_{1,n} \geq 0, \ n \in N_{k+1}, \\
& Z_{2,n} + q_{2,n} \geq (E_{k,n}^{max} - E_{k,n}^{min})/2 \times y_2, \ q_{2,n} \geq 0, \ Z_{2,n} \geq 0, \ n \in N_{k+1}, \\
& y_1 \geq 1, \ y_2 \geq 1, \ \eta \geq 0, \ a = \sum_{n=1}^{N_{k+1}} p_{k+1,n}, \\
& a\hat{u}_i + \tau_i^T (h - H\hat{u}_i) \leq \beta_i, \ \tau_i \geq 0, \ \|H^T \tau_i + a\|_\infty \leq \eta, \ i \in [S],
\end{aligned} \quad (8)$$

where $Z_{1,n}, q_{1,n}, y_1, Z_{2,n}, q_{2,n}, y_2, \tau_i, \beta_i, \eta$ are the auxiliary variables introduced in the model conversion. $\Gamma \in [0,1]$ is the robustness parameter, and the robustness is stronger as its value is larger; the impact of this on the evaluation results are discussed in the case studies. The conversion details can be found in [9][10].

The above MILP model can be efficiently solved via common solvers, e.g., CPLEX or Gurobi. Its scalability regarding the EV numbers is tested in the case studies.

## IV. Case Studies

Two types of electric vehicles are used for simulation to reflect the diversity of electric vehicle models, as shown in Table I, which presents common EV data used in many studies, such as [7]. The simulation is performed over 12 consecutive time intervals with 5-min resolutions to reflect the characteristics of online rolling evaluation. Historical data points of $u$ can be calculated through the open EV charging dataset [8]. In each interval, 20 historical data points of $u$ are used in the simulation; these points are between 0.8 and 1.0. The Gurobi solver in the MATLAB environment is adopted for modeling and simulation.

TABLE I. Electric Vehicle Information

| Type | 1 | 2 |
|---|---|---|
| Charge/discharge power(kW) | 40 | 60 |
| Battery capacity(kWh) | 60 | 82 |
| Battery capacity lower limit(kWh) | 5 | 10 |
| Charge/discharge efficiency | 0.95 | 0.95 |
| Quantity | 50 | 50 |

To verify the necessity of considering uncertain departure behavior and the SOC, we adopt the following four methods to evaluate EV flexibility. All the evaluation results are compared with the actual flexibility $[\underline{P}_k^{act}, \overline{P}_k^{act}]$ that is obtained *post hoc*, i.e., based on the actual SOC and departure time that is readily observed in the simulation.

*Method 1* (***M1***): Only considers uncertain departure behavior.
*Method 2* (***M2***): Only considers uncertain SOC.
*Method 3* (***M3***): Neglects to consider the uncertainties.
*Method 4* (***M4***): Proposed method considering both uncertainties.

The simulation results are shown in Table II and Fig. 2(a)-(d). Here, UBC denotes *upper-boundary-crossing* count, i.e., the number of times $\overline{P}_k$ is larger than $\overline{P}_k^{act}$. LBC denotes *lower-boundary-crossing* count, i.e., the number of times $\underline{P}_k$ is lower than $\underline{P}_k^{act}$. OEF denotes *over-evaluated flexibility*, e.g., the red area in Fig. 2. UEF denotes *under-evaluated flexibility*, e.g., the blue area in Fig. 2. Since the UBC, LBC and OEF of Methods M1-M3 are not 0, these results indicate that insufficient consideration of the uncertainties will lead to overestimation of the flexibility, which will introduce the risk of the aggregator not having enough flexibility to respond to the real-time regulation signal.

TABLE II. The Simulation Results of The Four Methods

| Methods | UBC | LBC | OEF (MW) | UEF (MW) |
|---|---|---|---|---|
| M1 | 0 | **4** | 1.96 | 6.58 |
| M2 | **10** | **2** | 2.30 | 5.96 |
| M3 | **10** | **9** | 6.53 | 0.73 |
| M4 (proposed) | 0 | 0 | 0 | 13.05 |

The sensitivity of $\Gamma$ is then tested. The results are shown in Table III and Fig. 2(d)-(f). As expected, the robustness of the evaluated results decreases as $\Gamma$ decreases. However, even if $\Gamma = 0.3$, the proposed method still performs better than the other three methods, which again verifies the advantage of using the proposed method over the others.

TABLE III. The Simulation Results for Different $\Gamma$

| Case | UBC | LBC | OEF (MW) | UEF (MW) |
|---|---|---|---|---|
| $\Gamma = 1.0$ | 0 | 0 | 0 | 13.05 |
| $\Gamma = 0.7$ | 0 | 1 | 0.24 | 9.75 |
| $\Gamma = 0.3$ | 0 | 3 | 1.20 | 7.68 |

We also test the solution time of the proposed method on a single interval with different EV numbers. The results are summarized in Table IV. These results verify the scalability of the proposed method regarding the EV numbers.

TABLE IV. Simulation Time Statistics With Different EV Numbers

| The EV numbers | 100 | 200 | 500 | 1000 |
|---|---|---|---|---|
| Time(s) | 1.11 | 1.24 | 1.44 | 1.75 |

Fig. 2. The comparison diagram of the flexibility evaluation results. (a) M1; (b) M2; (c) M3; (d) M4 with $\Gamma = 1.0$; (e) M4 with $\Gamma = 0.7$; (f) M4 with $\Gamma = 0.3$.

## V. Conclusion

In this letter, the impact of participating in the real-time market and regulation on EVs' SOC and the impact of random early departure is identified. A novel distributionally robust real-time flexibility evaluation method is proposed considering these uncertainties. The evaluation safety and the computational scalability are demonstrated through numerical tests. The proposed method can help the aggregator report reliable real-time flexibility, which benefits both the aggregator itself and the SO in the regulation service.